\documentclass[]{aa}
\usepackage{graphicx}
\usepackage{txfonts}
\def\fvdt#1 {{\bf#1}}

\begin{document}
\title{Recent Searches for the Radio Lines of NH$_3$ in Comets}

%\subtitle{none}

\author{
        J. Hatchell, \inst{1,3},
        M.K. Bird, \inst{2},
        F.F.S. van der Tak, \inst{1}
        \and
        W.A. Sherwood, \inst{1}
           }

   \institute{Max-Planck-Institut f\"ur Radioastronomie,  Auf dem H\"ugel 69,
             53121 Bonn, Germany
             \and
             Radioastronomisches Institut, Universit\"at Bonn,
              Auf dem H\"ugel 71, 53121 Bonn, Germany
              \and
              School of Physics, University of Exeter,
Stocker Road, Exeter EX4 4QL, UK; hatchell@astro.ex.ac.uk
             }

   \date{Received ; accepted }

   \abstract{
Radio observations in the ammonia inversion lines of four comets,
C/2001 A2 (LINEAR), 153P/Ikeya-Zhang, C/2001 Q4 (NEAT)
and C/2002~T7 (LINEAR),
were performed at the Effelsberg 100-m Radio Telescope during their
respective close approaches to Earth.
% With the exception of a possible marginal detection ($\sim2 \sigma$) in
% Comet Ikeya-Zhang,
None of the four lowest energy metastable lines
$(J,K=J),J=1\hbox{--}4$, could be detected in these comets.
We derive the following 3$\sigma$ upper bounds on the NH$_3$ production rate , and comparing to the corresponding water production rates, percentage NH$_3$ abundances relative to H$_2$O:
 $Q(\hbox{NH}_3) < $~1.9$\times$10$^{26}$~s$^{-1}$ (0.63\%) for C/2001 A2 (LINEAR),
$Q(\hbox{NH}_3) < $~2.7$\times$10$^{26}$~s$^{-1}$ (0.13\%) for C/2001 Q4 (NEAT),
$Q(\hbox{NH}_3) < $~2.3$\times$10$^{27}$~s$^{-1}$ (0.74\%) for C/2002 T7 (LINEAR) and
%Optimistically assuming a positive detection in Comet 153P/Ikeya-Zhang,
%the implied ammonia production rate is
$Q(\hbox{NH}_3) \leq $~6.3$\times$10$^{26}$~s$^{-1}$ (0.63\%)  for
Comet 153P/Ikeya-Zhang. At 0.74\% or less, the
ammonia-to-water ratios are factors of $\sim$2 below the value for C/1995 O1
(Hale-Bopp) and 1P/Halley, suggesting chemical diversity between
comets. The 18-cm lines of OH were clearly detected in the two comets
observed during the 2004 campaign, thereby validating the cometary
ephemerides.   }

   \titlerunning{Searches for radio lines of NH$_3$}
   \authorrunning{Hatchell et al.}
   \maketitle
%
%________________________________________________________________

\section{Introduction}

Attempts to detect the radio K-band lines of ammonia (NH$_3$) were
made during the recent apparitions of four comets with the 100-m
Effelsberg Radio Telescope of the Max-Planck-Institut f\"ur
Radioastronomie (MPIfR).  Observations were performed on Comets C/2001
A2 (LINEAR), 153P/Ikeya-Zhang, C/2001 Q4 (NEAT) and C/2002 T7
(LINEAR).  A preliminary account of the observations of comets
C/2001~A2 and 153P was previously published by Bird et al.~(2002).
Ammonia had previously been detected at MPIfR in the comets
C/1983~H1 (IRAS-Araki-Alcock) (Altenhoff et al.~1983) and C/1995~O1 (Hale-Bopp) (Bird et
al.~1997, 1999), and at the Green Bank Telescope in C/1996~B2 (Hyakutake)
(Palmer et al.~1996)\@.  Column density predictions from model
calculations were encouraging for these recent comets due to their
relatively favorable viewing geometries.

Ammonia is expected to be among the more abundant volatile parent
molecular constituents of cometary nuclei.
Chemical models predict an abundance relative to H$_2$O ice
  of a few percent (Charnley \&
Rodgers~\cite{charnleyrodgers02}). Our observational knowledge
about NH$_3$, ironically, comes largely from optical spectra of
NH$_2$, its longer-lived dissociation product.   A recent
reanalysis of the NH$_2$ spectra of many comets suggests a typical
NH$_3$ abundance of 0.5\% (Kawakita \& Watanabe
\cite{kawakitawatanabe02}, using the comet database of
Fink \& Hicks~\cite{finkhicks96}). 
% Fink et al. 1999 & earlier refs: database of many comets
% inc. Halley, C/1996 B2 Hyakutake, 81 P/Wild 2, 46 P/ Wirtanen

This is less than the estimates of 1.0-1.8\% from the radio detections
of NH$_3$ in C/1995~O1 (Hale-Bopp) (Bird et al.~\cite{bird97,bird02};
Hirota et al. \cite{hirota99}), but consistent with the 
revised estimate of 0.6\% in C/1996~B2 (Hyakutake)
(Bird et al.~ \cite{bird99}).
The NH$_2$ data 
imply an ammonia abundance of 0.75\% for comet 1P/Halley,
i.e.~only about half the {\it in situ} value
of 1.5\% NH$_3$ measured during the flyby of 
the Giotto spacecraft (Meier et al.~\cite{meier94}).
 
In the interstellar
medium, for comparison, the best estimates we have for the NH$_3$
abundance in ice  are upper limits of
5\% and 7\% on the lines of sight towards W33A (Taban et al.~\cite{taban03})
and the massive protostars GL 989 and GL~2136
(Dartois et al.~\cite{dartois02}), respectively.

Few radio spectra of NH$_3$ have been obtained because of the short
photodissociation lifetime in the interplanary medium near 1~AU ($\tau
\simeq$~6000~s).  The diameter of the ammonia cloud around comets is
typically only a few thousand kilometers and beam dilution can be
significant.  The radio line observations at MPIfR were thus scheduled
near the relatively close approaches of the comets to Earth, thereby
insuring that the source region of emission would be larger than, or
at least only slightly smaller than, the $40''$ MPIfR beam.  Contrary
to expectations, no NH$_3$ lines could be detected with certainty
during the MPIfR comet observation campaigns in 2001-2004.  The
nondetections in C/2001~A2, 153P, C/2001~Q4 and C/2002~T7 suggest an
underabundance of NH$_3$ in these comets with respect to Comet C/1995
O1 (Hale-Bopp) and 1P/Halley.

%%%%%%%%%%%%%%%%%%%%%%%%%%%%%%%%%%%%%%%%%%%%%%%%%%%%%%%%%%%%%%%%
%                             Table obs_elem_table
%%%%%%%%%%%%%%%%%%%%%%%%%%%%%%%%%%%%%%%%%%%%%%%%%%%%%%%%%%%%%%%%%
\begin{table*}
\begin{minipage}[t]{\columnwidth}
\caption[]{Orbital Elements: MPIfR Comet Observation Campaign 2001-2004}
\label{obs_elem_table}
\centering
\renewcommand{\footnoterule}{}  % to avoid a line before footnotes
\begin{tabular}{lrrrr}
\hline 
 & &  \\[-3mm]
 Comet    & C/2001 A2-B & 153P     &  C/2001 Q4 &  C/2002 T7  \\[1mm]
          & (LINEAR)    & (Ikeya-Zhang) &  (NEAT)    &   (LINEAR) \\[1mm]
\hline 
 & &  \\[-3mm]
%Orb.~element set$^{\, a}$  & 2001-M48 & 2002-H05  &  2004-J69 &  2004-J05  \\
Orb.~element set~\footnote{\small Minor Planet Electronic Circular (MPEC) designation} & 2001-M48 & 2002-H05
& 2004-J69 & 2004-J05  \\
Asc.~node $\Omega$ ($\circ$) & 295.12564 & 93.37048 & 210.27852  &  94.85882 \\
Arg.~perih.\ $\omega$ ($\circ$) &295.32853 & 93.37048 & 1.20653 & 157.73671 \\
Inclination $i$ ($\circ$)   &36.47538 & 28.12159  & 99.64258    & 160.658324  \\
Perih.~dist.\ $q$ (AU)    &0.7790280 & 0.5070583  & 0.9619575   & 0.6145967  \\
Eccentricity $e$     & 0.9993455 & 0.9899505   & 1.0007438     & 1.0005157  \\
Perih.~epoch     & 2001 May 24.52062 & 2002 Mar 18.97927 
& 2004 May 15.96707 & 2004 Apr 23.06172   \\
Osc.~epoch~\footnote{\small Epoch of osculating elements}       & 2001 May 11.0 & 2002 Mar 27.0
& 2004 May 8.0 & 2004 Jun 4.0  \\
\hline 
\end{tabular}

\end{minipage}
\end{table*}

% New table May 17th from Mike Bird's email - updated int. times
%%%%%%%%%%%%%%%%%%%%%%%%%%%%%%%%%%%%%%%%%%%%%%%%%%%%%%%%%%%%%%%%
%                             Table obs_param_table
%%%%%%%%%%%%%%%%%%%%%%%%%%%%%%%%%%%%%%%%%%%%%%%%%%%%%%%%%%%%%%%%%
\begin{table*}
\caption[]{Mean Observation Parameters:
MPIfR Comet Observation Campaign 2001-2004} 
\label{obs_param_table}
\centering
\renewcommand{\footnoterule}{}  % to avoid a line before footnotes
\begin{tabular}{lccrrccccccc}
\hline 
 & &  \\[-3mm]
Comet  & \multicolumn{1}{c}{$\Delta t\,$\footnote{Integration time on comet}}
& \multicolumn{1}{c}{elevation} 
& \multicolumn{2}{c}{RA\_(J2000)\_DEC}
& \multicolumn{1}{c}{$\Delta$} & \multicolumn{1}{c}{$r$}
& \multicolumn{1}{c}{$\angle$SEC~\footnote{Sun-Earth-Comet,
solar elongation angle}}
& \multicolumn{1}{c}{$\angle$SCE~\footnote{Sun-Comet-Earth, phase angle}}
& $d\,$\footnote{Estimated NH$_3$ coma diameter} 
& $\Delta \theta\,$\footnote{MPIfR beam diameter (3 dB) at comet} 
& \underline{~~$d$}  \\
~~date  & \multicolumn{1}{c}{(min)} 
&  \multicolumn{1}{c}{($\circ$)} & (hr mn) & ($\circ$ $\prime$) &
\multicolumn{1}{c}{(AU)} & \multicolumn{1}{c}{(AU)}
& \multicolumn{1}{c}{($\circ$)} & \multicolumn{1}{c}{($\circ$)}
& (10$^3$ km) & (10$^3$ km) & $\Delta \theta$ \\[1mm]
\hline 
 & &  \\[-3mm]
\multicolumn{7}{l}{\underline{C/2001 A2-B (LINEAR)} in 2001}\\
~~07.22 Jul & 70 & 43 & 23 31 & 05 37 & 0.262 & 1.130 & 109 & 58 & 9.7 & 7.9 & 1.22 \\
\multicolumn{7}{l}{\underline{C/2002 C1 (Ikeya-Zhang)} in 2002}\\
~~25.42 Apr & 53 & 57 & 20 50 & 61 29 & 0.409 & 0.974 & 74 & 83 & 9.5 & 12.3 &
0.77 \\
~~04.40 May & 83 & 58 & 18 19 & 55 30 & 0.410 & 1.124 & 95 & 63 & 11.8 & 12.3 &
0.96 \\
~~06.38 May & 17 & 51 & 17 55 & 52 38 & 0.416 & 1.157 & 100 & 59 & 12.3 & 12.5 &
0.98 \\
~~07.41 May & 45 & 54 & 17 43 & 51 14 & 0.421 & 1.174 & 102 & 57 & 12.6 & 12.7 &
0.99\\
\multicolumn{3}{l}{\underline{C/2001 Q4 (NEAT)} in 2004}\\
~~08.73 May & 47 & 29 & 07 44 & --07 53 & 0.327 & 0.970 & 74 & 87 & 9.3 & 9.9 & 0.94 \\
~~13.69 May  & 90 & 53 & 08 23 & +13 54 & 0.390 & 0.963 & 72 & 86 & 9.1 & 11.7 & 0.78 \\
~~17.65 May  & 82 & 65 & 08 47 & +25 42 & 0.473 & 0.962 & 70 & 82 & 9.1 & 14.3 & 0.64 \\
\multicolumn{3}{l}{\underline{C/2002 T7 (LINEAR)} in 2004}\\
~~08.40 May & 35 & 35 & 00 41 & $-$4 06 & 0.534 & 0.699 & 41 & 109 & 5.4 & 16.1 & 0.34 \\
~~13.44 May & 63 & 28 & 01 43 & $-$10 06 & 0.364 & 0.755 & 38 & 125 & 6.0 & 11.0 & 0.55 \\
 & &  \\[-3mm]
\hline 
\end{tabular}
\end{table*}

\section{Ammonia Line Observations}
\label{sect:obs}

%============================ figure 1 =============================
   \begin{figure*}
   \centering
   \includegraphics[width=\textwidth]{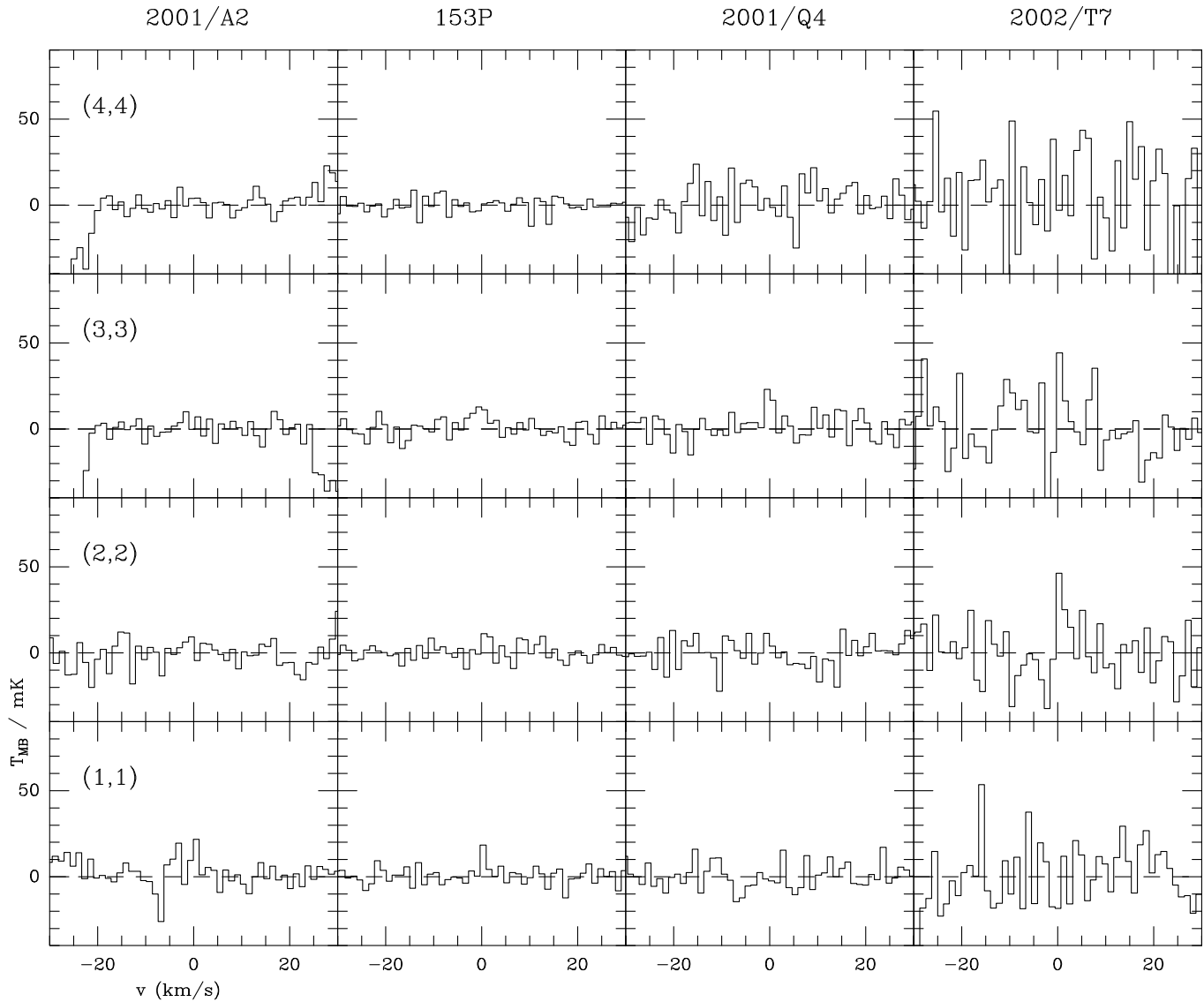}
      \caption{NH$_3$ spectra of all comets (left to right: C/2001~A2, 153P, C/2001~Q4, C/2002~T7) in
the lowest 4 metastable levels, 
centered at zero velocity in the comet rest frame.
              }
         \label{plotallcomets}
   \end{figure*}

The 18-26 GHz HEMT receiver system (primary focus) was used for this
program.  The lowest four inversion transitions of ammonia in its
metastable states $(J,K=J),\,J=1\hbox{--}4$, were observed
simultaneously in split-mode, dividing the autocorrelator into 4 bands
with 2048 channels each.  Raw spectra were recorded at a typical
channel spacing of 0.25~km~s$^{-1}$ over the range of at least
$\pm$30~km~s$^{-1}$ about the expected line frequency at zero velocity
in the comet rest frame.  Spectra were calibrated using W3(OH) and
system temperatures were generally between 35 and 60 K\@.  Various
observing modes were used: frequency switching (C/2001~A2),
position switching every 30s with a throw of either $5'$ or
$10'$ (C/2001~A2, 153P), and beam switching using the rotating horn
with a beam throw of $2'$ and a 1s cycle (C/2001~Q4, C/2002~T7).
Table~\ref{obs_elem_table} presents the orbital elements used to track
the comet position and velocity.

 A summary of the mean geometric parameters during the observations,
  including integration time $\Delta t$, geocentric distance $\Delta$,
  and heliocentric distance $r$, is given in
  Table~\ref{obs_param_table}.  This table also shows values for the
  estimated diameter of the NH$_3$ cloud, $d = \delta v \cdot \tau
  \cdot r^2$, with $r$ the heliocentric distance (normalized to 1~AU),
  $\tau$ the mean lifetime of an ammonia molecule in interplanetary
  space, and $\delta v$ the velocity spread of the coma gas assuming
  spherically symmetric outflow.  The mean lifetime of an ammonia
  molecule in interplanetary space depends on the solar UV flux and is
  probably somewhat longer than average during the 2004 epoch near
  solar minimum.  In the following we use values for the
  photodissociation lifetime of NH$_3$ at 1~AU in the range 5400 s$<
  \tau <$ 5800~s.  Different values for the line width were
  taken for each individual comet: 1.4 km/s (C/2001 A2);
  1.8 km/s (153P); 1.7 km/s (C/2001 Q4); and 1.9 km/s (C/2002
  T7).  These were selected on the basis of independent radio line
observations of other molecules during the same epoch (Biver et al.,
in preparation).  The cloud diameter $d$ may be compared to the next
column, the physical extent of the FWHM antenna beam at MPIfR
($\Delta \theta$, with $\theta = 40^{\prime \prime}$) at the comet.
The ratio of these quantities, $d/\Delta \theta$, is shown in the last
column of Table ~\ref{obs_param_table}.  As a rule, the detection
probablility is more favorable for large values of this ratio.  The
best conditions for this criterion held for comet C/2001 A2 and were
still fairly good for comet 153P\@.  The beam was distinctly
larger than the ammonia cloud for most of the observations in 2004.

\begin{table*}
\begin{minipage}[t]{\columnwidth}
\caption[]{Ammonia line observations of comets at MPIfR and
upper limits on the NH$_3$ column density.
% [COLUMN 6 WAS RECALCULATED; FOOTNOTES WERE REARRANGED;
% $\Delta v$ IS NOW $\delta v$;
% COMET P/153: COLUMN 5 WAS RECALCULATED AND
% $\delta v$ = 1.9 IS NOW $\delta v$ = 1.7.]
}
\label{obs_table}
\centering
\renewcommand{\footnoterule}{}  % to avoid a line before footnotes
\begin{tabular}{cccccccc}
\hline
line & frequency &
$T_{rms}$ &
$T_{99.7\%}$ &
%$[ \int T_{\scriptscriptstyle{MB}} \; dv ]_{max}^{\; b}$ &
$[ \int T_{\scriptscriptstyle{MB}} \; dv ]_{max}\,$\footnote[1]{$1.064 \cdot T_{99.7\%} \cdot \delta v$ with $\delta v$ values as given in the table.} &
$\langle N(J,J) \rangle_{max}\,$\footnote[2]{from Eq.~(\ref{NJJ})}  
&$n(J,J)$\footnote[3]{defined in Eq.~(\ref{njj})}
&$\langle N(\hbox{NH}_3) \rangle_{max}$
\\
(J,J)    & &  [mK] &  [mK] 
& [mK$\cdot$km s$^{-1}$]
& [10$^{12}$ cm$^{-2}$]
&  
& [10$^{12}$ cm$^{-2}$]
\\ 
\hline
  & & &  & & &&\\[-3mm]
\multicolumn{6}{l}{Comet C/2001 A2-B (LINEAR):
7 Jul 2001 ($\Delta t$~= 70 min, $\delta v = 1.4\hbox{ km s}^{-1}$)}  \\
(1,1) & 23.69450 &   16 & 36 & 54 & 0.73 &0.26 &$<2.8$\\
(2,2) & 23.72263 &   16 & 30 & 45 & 0.46 &0.18 &$<2.5$\\
(3,3) & 23.87013 &   16 & 34 & 51 & 0.46 &0.20 &$<2.3$\\
(4,4) & 24.13942 &   15 & 30 & 45 & 0.38 &0.05 &$<7.6$\\
\multicolumn{6}{l}{Lowest upper bound
\footnote[4]{Lowest upper limit on $\langle N(\hbox{NH}_3) \rangle$
from the four lines observed} } &     &\boldmath$<2.3$ \\
  & & &  & & &&\\[-3mm]
\multicolumn{6}{l}{Comet 153P/Ikeya-Zhang:
% 25 Apr, 4,6,7 May 2002 ($\Delta t$~= 53 min, $\delta v = 1.8\hbox{ km s}^{-1}$)}  \\
% (1,1) & 23.69450 &   12 & 30 %40 & 57 & 0.85 &0.26 & $<3.0$\\
% (2,2) & 23.72263 &   15 & 28 %26 & 54 & 0.47 &0.18 & $<3.0$\\
% (3,3) & 23.87013 &   16 & 40 %40 & 77 & 0.61 &0.20 & $<3.5$\\
% (4,4) & 24.13942 &   14 & 16 %14 & 34 & 0.15 &0.05 & $<5.9$\\
% Total &          &      &    &    &      &     &\boldmath$<3.0$ $^d$\\
25 Apr 2002 \footnote[5]{Calculations based on the best single epoch of
observations} ($\Delta t$~= 53 min, $\delta v = 1.8\hbox{ km s}^{-1}$)}  \\
(1,1) & 23.69450 &   12 & 40  & 77 & 1.04 &0.26 & $<4.0$\\%40
(2,2) & 23.72263 &   15 & 26  & 50 & 0.51 &0.18 & $<2.8$\\%26
(3,3) & 23.87013 &   16 & 40  & 77 & 0.69 &0.20 & $<3.5$\\%40
(4,4) & 24.13942 &   14 & 14  & 27 & 0.23 &0.05 & $<4.6$\\%14
\multicolumn{6}{l}{Lowest upper bound $^d$} &     &\boldmath$<2.8$ \\
  & & &  & & &&\\[-3mm]                       
\multicolumn{6}{l}{Comet C/2001 Q4 (NEAT):
8,13,17 May 2004 ($\Delta t$~= 219 min, $\delta v = 1.7\hbox{ km s}^{-1}$)}  \\
(1,1) & 23.69450 &  18 & 20 & 36 & 0.49 &0.26 & $<1.9 $\\
(2,2) & 23.72263 &  19 & 28 & 51 & 0.52 &0.18 & $<2.9 $\\
(3,3) & 23.87013 &  19 & 44 & 80 & 0.72 &0.20 & $<3.6 $\\
(4,4) & 24.13942 &  23 & 20 & 36 & 0.31 &0.05 & $<6.2$\\
\multicolumn{6}{l}{Lowest upper bound $^d$} &     &\boldmath$<1.9$ \\
  & & &  & & &&\\[-3mm]
\multicolumn{6}{l}{Comet C/2002 T7 (LINEAR): 
8,13 May 2004 ($\Delta t$~= 98 min, $\delta v = 1.9\hbox{ km s}^{-1}$)}  \\
(1,1) & 23.69450 &  29 & 30  & 61 & 0.82 &0.26 & $<3.2 $\\
(2,2) & 23.72263 &  29 & 82  & 166 & 1.69 &0.18 & $<9.4 $\\
(3,3) & 23.87013 &  29 & 52  & 105 & 0.95 &0.20 & $<4.8 $\\
(4,4) & 24.13942 &  47 & 80  & 162 & 1.37 &0.05 & $<27.4$\\
\multicolumn{6}{l}{Lowest upper bound $^d$} &     &\boldmath$<3.2$ \\
\hline
\end{tabular}

\end{minipage}
\end{table*}

The Comet C/2001~A2 was observed to split into multiple nuclei
(Sekanina et al., 2002).
The position of the largest fragment (denoted `A2-B'), targeted
for these observations, was derived from the
then current ephemeris (orbital elements in MPEC 2001-M48).
The initial observations of Comet 153P on 25 Apr 2002, 
a long session in good weather, hinted at
a marginal detection in the (1,1) and (3,3) lines (see Bird et al., 2002)\@.
Follow-up observations performed on 4, 6 and 7 May under less favorable
weather conditions (and higher system noise temperatures), however,
could not confirm the tentative ammonia lines.  
Comet C/2001~Q4 could not be observed at MPIfR until early May~2004
due to its high southern declination.
The southward motion of Comet C/2002~T7 became problematic as it approached the Earth,
thereby severely restricting the last observation interval.
In fact, no NH$_3$ observations of Comet C/2002~T7 could be performed on 17~May~2004
due to lack of time following the OH observations (see next section).
This and the unfavorable declination resulted
in the shorter accumulated integration time.

The summed spectra for each ammonia line from all comet observations
are displayed on the same scale 
($T_{\scriptscriptstyle{MB}}$ in mK over $\pm$30~km/s)
in Fig.~\ref{plotallcomets}.
The four spectra (from bottom to top) correspond to the
(1,1) to (4,4) metastable states,
and are centered at zero velocity in the comet rest frame.

  Upper limits to the line amplitudes are estimated from the spectra
  by fitting a Gaussian line model.  This is preferable to taking the
  3$\sigma$ upper limit from the RMS noise level calculated over the
  whole spectrum because the actual data values in the central
  channels are taken into account.  Therefore if there is any evidence for a
  weak line, that will increase our upper limits on the line
  amplitudes accordingly, in order to definitely exclude the weak
  line.  This was the case for Comet 153P/ Ikeya-Zhang.
  
  The model spectrum consists of a Gaussian line of amplitude $A$ plus
  a baseline level $B$.  The baseline level is considered again here
  as it is critical to the final line amplitudes, although linear
  baselines have already been subtracted from the spectra. 
 The linewidth and line centre are fixed:
  the line centre always at 0~km~s$^{-1}$,
and the linewidth taken from the observed
  radio linewidths of other molecules (Biver et al., in preparation; see Table~\ref{obs_table}.
  A grid in parameter space is
  calculated for the line amplitude $A$ and the baseline level $B$.
  We calculate the probability $P(A,B)$ that the line has an amplitude
  $A$ with baseline $B$ using a likelihood function $\exp(-\chi^2/2)$
  with the constraint that $A$ must be positive.  As we are not
  interested in any residual baseline $B$, we integrate over $B$ for
  each value of $A$ to calculate the probability of each amplitude
  independent of baseline, $P(A) = \int P(A,B)\, dB \simeq \sum_{B_j}
  P(A_i, B_j)$\@.  From $P(A)$ we identify the 99.7\% upper limit on $A$
  (equivalent to 3$\sigma$ for a normal distribution). 
 The corresponding upper limit on the line integrated intensity
is calculated from $1.064\cdot A \cdot \delta v$\@. 
A summary of the NH$_3$ observations and derived results for all comets is
given in Table~\ref{obs_table}.

% The upper bounds on the line strengths were derived from the (3$\sigma$)
% baseline noise temperature $T_{3\sigma} = 3 T_{rms}$, assuming a Gaussian
% cometary emission line with a typical halfwidth of $\delta v$~=~1.8~km/s.
%"[LAST PARAGRAPH MOVED TO SECTION 4]"

\section{Hydroxyl Line Observations}

%
%%%%%%%%%%%%%%%%%%%%%%%%%%%%%%%%%%%%%%%%%%%%%%%%%%%%%%%%%%%%%%%%
%                             Table OH_obs_table
%%%%%%%%%%%%%%%%%%%%%%%%%%%%%%%%%%%%%%%%%%%%%%%%%%%%%%%%%%%%%%%%%
\begin{table*}
\begin{minipage}[t]{\columnwidth}
\caption[]{OH Line Observations of Comets at MPIfR, 17 May 2004}
\label{OH_obs_table}
\centering
\renewcommand{\footnoterule}{}  % to avoid a line before footnotes
\begin{tabular}{cccccc}
\hline
transition & frequency &  $T_{peak}$ & $v_0$ & $FWHM = \delta v$ &
Line Area~\footnote[1]{$\int T_{\scriptscriptstyle{MB}}\; dv$
(MPIfR gain: 2.0 K/Jy at 1665/1667 MHz)}  \\
$F_u \rightarrow F_l$  & [MHz] &  [mK] & [km s$^{-1}$] & [km s$^{-1}$] &
[mJy$\cdot$km s$^{-1}$] \\
\hline
  & & & &   & \\[-3mm]
\multicolumn{6}{l}{Comet C/2001 Q4 (NEAT): 17 May 2004 ($\Delta t$~=~12 min)}  \\ 
1 $\rightarrow$ 1 & 1665.4018 &  72~\footnote[2]{$T_{3\sigma}$} & -- & -- & --   \\
2 $\rightarrow$ 2 & 1667.3590 &  119 & -0.16$\pm$0.26 & 2.06$\pm$0.55 &
130$\pm$31  \\
  & & &  & & \\[-3mm]
\multicolumn{6}{l}{Comet C/2002 T7 (LINEAR): 17 May 2004 ($\Delta t$~=~16 min)} \\
1 $\rightarrow$ 1 & 1665.4018 & 99 & 0.54$\pm$0.23 & 2.69$\pm$0.36 &
142$\pm$23    \\
2 $\rightarrow$ 2 & 1667.3590 & 175 & 0.44$\pm$0.22 & 2.18$\pm$0.47 &
203$\pm$39  \\
\hline
\end{tabular}
\end{minipage}
\end{table*}

\begin{figure}
   \centering
   \includegraphics[width=\columnwidth]{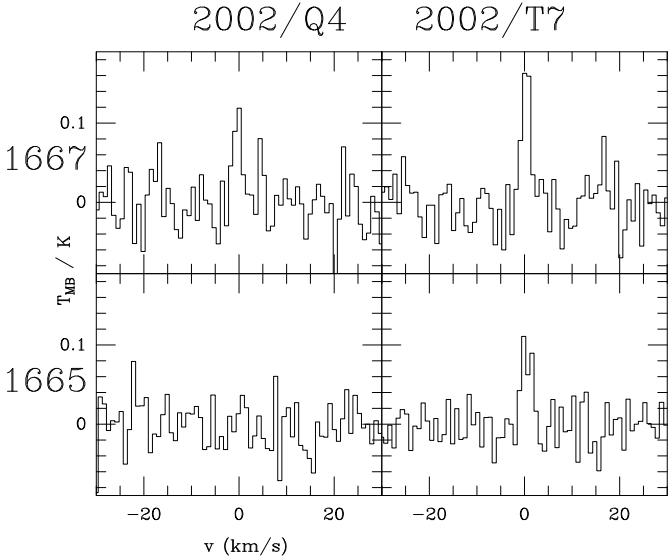}
\caption{ OH radio line spectra 
recorded on 17 May 2004. 
Left: C/2001 Q4 (NEAT), 12 minutes integration time;
Right: C/2002 T7 (LINEAR), 16 minutes integration time.
The main beam brightness
temperature $T_{\scriptscriptstyle{MB}}$ is shown for
the two OH lines at 1667~MHz (upper) and 1665~MHz (lower).
}
\label{plotoh}
\end{figure}

Following the nondetection of NH$_3$ lines during the 
observing sessions of Comets C/2001~Q4 and C/2002~T7 on 8 and 13 May 2004,
it was decided to verify the telescope comet
tracking configuration and, at the same time,
the accuracy of the ephemerides.
An imprompu change of receiver was requested and the initial hours of
the final allocated observation interval on 17 May were devoted to observing  
the two strongest hyperfine transitions of the
$\Lambda$-doublet ground-state of OH\@. 
The OH spectra recorded for each comet are shown in Fig.~\ref{plotoh}.
The upper (lower) panels show the two strongest OH lines at 1667 (1665) MHz,
centered at zero velocity in the rest frame of the comet ephemeris.
The total integration times, for comparison with the NH$_3$ observations,
were only 12 and 16 minutes on the Comets C/2001~Q4 and C/2002~T7, respectively.

A tabular summary of these observations, which
were the first cometary OH-observations performed at MPIfR
since Comet Halley (Bird et al.~1986),
is given in Table~\ref{OH_obs_table}.
The intensity of the $\Lambda$-doublet lines is governed to a large
extent by the inversion parameter, a measure of the imbalance in the
upper and lower levels of the OH ground state (Schleicher and A'Hearn,
1988).  If the two levels are nearly equal, the inversion parameter is
zero and the OH-maser is deactivated.  Although this was very nearly
the case for both comets on 17 May 2004, clear detections were
obtained for all combinations except for comet C/2001~Q4 at 1665 MHz.
The 1667 MHz line is about twice as strong as the 1665 MHz line
in both comets, consistent with the 9:5 ratio expected under LTE
conditions.
%fvdt: Made last statement a bit more quantitative

It was verified that the OH 1667~MHz line strengths were consistent with
simultaneous measurements of the radio OH lines taken at the
Nan\c{c}ay Radio Telescope (J. Crovisier, private communication, see
Table~\ref{OH_compare}).  Note that the Nan\c{c}ay line strengths
given in the table represent an average over the 1665 and 1667~MHz lines,
both polarizations (LCP+RCP), and then scaled to the 1667~MHz line,
assuming they conform to the LTE ratio.  The line strengths at MPIfR
were found to be somewhat higher than at Nan\c{c}ay, but are within
the 1$\sigma$ measurement errors.   An exact agreement would
not be expected as although the Nan\c{c}ay and MPIfR beams at 18cm
are similar in area, they are very different shapes and therefore
couple to the comet OH emission differently (the Nan\c{c}ay beam is
elongated with FWHP $3.5' \times 19'$ , whereas the MPIfR beam is
approximately round with a FWHP $7.8'$).  

%The fact that the line strengths do agree so well suggests that the diameter of the OH emission is less than $3.5'$.\

%============================ figure 2 =============================

%%%%%%%%%%%%%%%%%%%%%%%%%%%%%%%%%%%%%%%%%%%%%%%%%%%%%%%%%%%%%%%%
%                             Table OH_compare
%%%%%%%%%%%%%%%%%%%%%%%%%%%%%%%%%%%%%%%%%%%%%%%%%%%%%%%%%%%%%%%%%
\begin{table}
\begin{minipage}[t]{\columnwidth}
\caption[]{OH Comet Observations, Nan\c{c}ay RT and MPIfR, May 2004}
\label{OH_compare}
\centering
\renewcommand{\footnoterule}{}  % to avoid a line before footnotes
\begin{tabular}{cccc}
\hline
May Date &  Line Area~\footnote[1]{Nan\c{c}ay RT, mean of 1665~MHz and 1667~MHz lines assuming LTE ratio of 5:9} &  Line Area~\footnote[2]{MPIfR, mean of 1665~MHz and 1667~MHz lines assuming LTE ratio of 5:9.  Limits for C/2001~Q4 are assuming the 1665~MHz line amplitude is (a) 0 (b) $3\sigma$ upper limit.} & Line Area~\footnote[3]{MPIfR, 1667~MHz detection only}  \\
&LTE &LTE &1667 only\\
 & [mJy$\cdot$km s$^{-1}$] & [mJy$\cdot$km s$^{-1}$]& [mJy$\cdot$km s$^{-1}$] \\
\hline
 & &   \\[-3mm]
\multicolumn{4}{l}{Comet C/2001 Q4 (NEAT)}  \\ 
16 &  104$\pm$12 & no obs. & no obs.\\
17 &  101$\pm$12 & 65--207 &  130$\pm$31  \\
18 &  75$\pm$12  & no obs. & no obs.\\
  & &  \\[-3mm]
\multicolumn{4}{l}{Comet C/2002 T7 (LINEAR)}  \\
16 &  176$\pm$12 & no obs. & no obs.\\
17 &  no obs.    & 229$\pm28$      & 203$\pm$39  \\
18 &  185$\pm$16 & no obs. & no obs.\\
\hline
\end{tabular}

\end{minipage}
\end{table}

\section{Upper limits on NH$_3$ column densities and production rates}
\label{sect:limits}
%
%%%%%%%%%%%%%%%%%%%%%%%%%%%%%%%%%%%%%%%%%%%%%%%%%%%%%%%%%%%%%%%%%
%                             Table model_param
%%%%%%%%%%%%%%%%%%%%%%%%%%%%%%%%%%%%%%%%%%%%%%%%%%%%%%%%%%%%%%%%%
\begin{table*}
\caption[]{Comet Production Model Parameters}
\label{model_param}
\vspace*{1mm}
\hspace*{8mm}
\begin{tabular}{lccccccl}
\hline
Comet                   & mm yyyy    & $Q_o$ & $r_o$ & $\delta v_o$ & $\alpha$  & $\tau$ & Reference \\
          &   & [10$^{29}$~s$^{-1}$] & [AU]     & [km s$^{-1}$]   & [s]    & \\
\hline
%C/1983 H1 (IAA)$^{\, a}$& 05 1983 & 0.33  & 1.005 & 1.8 & 3.0  & 5700 & Feldman et al.~(1983)\\
%1P/Halley              & 11 1985 & 6.9  & 0.903 & 1.8 & 3.0  & 5800 & Krankowsky et al.~(1986)\\
%C/1996 B2 (Hyakutake)  & 03 1996 & 1.7  & 1.090 & 1.8 & 3.0  & 5800 & Mumma et al.~(1996)\\
C/1995 O1 (Hale-Bopp)  & 04 1997 & 100  & 0.914 & 1.8 & 3.0  & 5600 & Colom et al.~(1999)\\
C/2001 A2-B (LINEAR)   & 07 2001 & 0.28 & 1.130 & 1.49 & 3.0  & 5400 & F. Bensch (priv.~comm.)\\
153P/Ikeya-Zhang       & 04 2002 & 0.92 & 1.000 & 1.77 & 3.21 & 5600 & Dello Russo et al.~(2004)\\
C/2001 Q4 (NEAT)       & 05 2004 & 1.8  & 1.017 & 1.67 & 3.0  & 5800 & N. Biver (priv.~comm.)\\
C/2002 T7 (LINEAR)     & 05 2004 & 4.5  & 0.645 & 1.59 & 3.0  & 5800 & N. Biver (priv.~comm.)\\
\hline

\end{tabular}
%\hspace*{4mm} $^{\, a}$ IRAS-Araki-Alcock\\
\end{table*}

%"[MOVED FROM END OF SECTION 2]"
Table 3 (column 6) shows the upper bounds on the column density
$\langle N(J,J) \rangle_{max}$ (in cm$^{-2}$) of molecules in the
given state.   These are calculated from the line integrated intensity
in K$\cdot$km s$^{-1}$
using the expression (e.g., Rohlfs \& Wilson 1996, p.~362)
\begin{equation}
\langle N(J,J) \rangle = 6.8 \times 10^{12} \frac{J+1}{J} 
\left[ \int{T_{\scriptscriptstyle{MB}}(v) dv} \right] 
\label{NJJ}
\end{equation}
Both the upper and lower levels of the specific metastable NH$_3$ state
are included in Eq.~(\ref{NJJ}).

Knowing the mean column density of the NH$_3$ molecules in a given metastable
state, the total column density of all NH$_3$ molecules is obtained from
the relation
\begin{equation}
n(J,J) = \langle N(J,J) \rangle / \langle N(\hbox{NH}_3) \rangle 
\label{njj}
\end{equation}
where $n(J,J)$, the relative population for each observed state,
depends on the mean kinetic temperature and density of the cometary
NH$_3$ gas in the antenna beam.  The population will favor the
metastable (K=J) states, particularly if the collision time is much
longer than the decay time scales $\sim$10 s for the nonmetastable
states.  We have performed statistical equilibrium calculations using
collision rate coefficients of ammonia in an H$_2$ background gas
(Danby et al.~\cite{danby88},
 Sch\"oier et al.~\cite{schoier05}),
but multiplied by a factor 4.3
to account roughly for the increase in cross section for collisions
with H$_2$O.  The larger mass of H$_2$O than H$_2$, which decreases
the collision velocity, was not taken into account.  The calculations
included the lowest 17 states for ortho NH$_3$ up to (J.K)= (6,0),
599~K above ground, and the lowest 24 states for para NH$_3$ up to
(5,1), 423~K above ground.  The population of higher energy states was
found to be negligible for densities up to 10$^9$ cm$^{-3}$ and
kinetic temperatures up to 200 K.  The ortho-to-para ratio was assumed
to be unity, but may, in fact, be slightly higher (Kawakita et al.,
2001).  The coma gas density for the observed comets, based on their
later determined production rates, lie in the range from 10$^5$ to
10$^6$ cm$^{-3}$ at a distance $R$~=~5000~km from the nucleus.  This
can be taken as a typical mean density of the background gas in the
telescope beam.  The model calculations show that the NH$_3$ partition
function in the coma is fairly insensitive to the actual density and
only moderately dependent on the kinetic temperature over the range of
plausible values from 50~K~$\le T_{k} \le$~100~K\@.  The calculated
values of $n(J,J)$ in statistical equilibrium with $T_{k}$~=~100~K are
shown in Table~\ref{obs_table} for each transition.  The remaining
column of Table~\ref{obs_table} presents the total NH$_3$ column
densities, $\langle N(\hbox{NH}_3) \rangle_{max}$, calculated from
Eq.~(\ref{njj}) for the upper bounds of each transition and comet
given in Table~\ref{obs_table}.   To calculate the
  production rates, we use the strongest upper limits on
  $N(\hbox{NH}_3)$ from among the four individual NH$_3$ lines.
\begin{figure*}
   \centering
   \includegraphics[width=\textwidth]{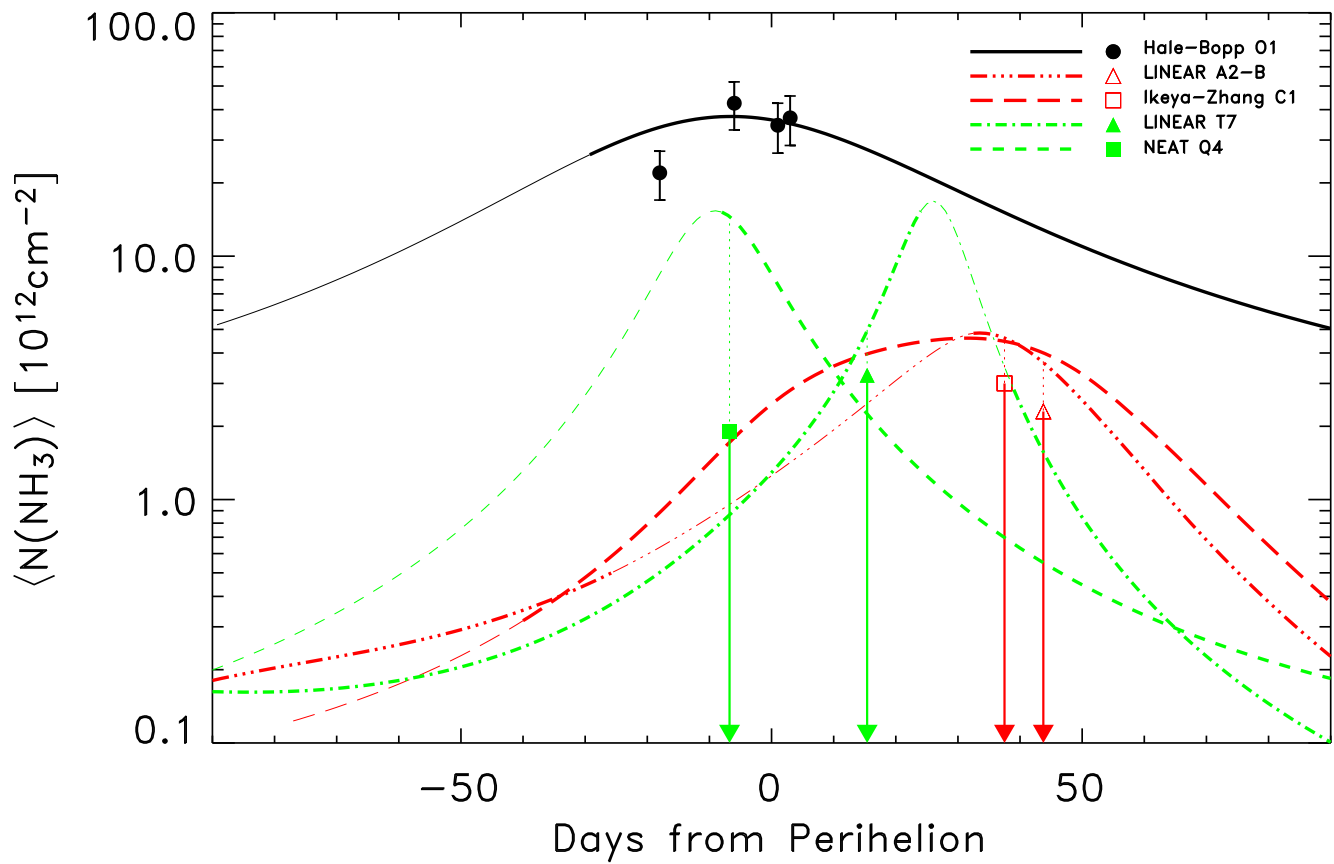}
\caption{Predicted and observed MPIfR beam-averaged NH$_3$
column densities for five comets.
The thick part of the curves denote intervals where the declination enables
observation at MPIfR ($\delta > -25^{\circ}$).
Maxima occur near closest approach to Earth. 
Points with standard deviations represent the line detections
in C/1995~O1 (Hale-Bopp) (Bird et al.~1997);
points without error bars are nondetection upper bounds for
the Comets C/2001~A2-B, 153P, C/2001~Q4 and C/2002~T7.
In all nondetection cases,
the upper bounds are significantly less than the predictions
based on an ammonia abundance of 1\%.  
The shortfall is indicated in each case by the dotted line.
}
\label{rcomet_log_2004} 
\end{figure*}

The upper bounds on the column densities for all observed comets are plotted in
Fig.~\ref{rcomet_log_2004} and compared
with pre-perihelion model predictions of $\langle N(\hbox{NH}_3) \rangle$
over an interval $\pm$90 days about perihelion (curves).
Fig.~\ref{rcomet_log_2004} also shows the measurements and prediction curve
for the firm detection of NH$_3$
in Comet C/1995~O1 (Hale-Bopp) (solid black line and filled circles).
% The predictions and upper bounds for the other four comets are as follows:
% %Comet A2-B (red dash-dotted line and open triangle);
% %Comet C1 (red long-dashed line and open square);
% %Comet C/2002~T7 (green dash-dotted line and solid triangle);
% %Comet C/2001~Q4 (green dashed line and solid square).
% Comet A2-B (dash-three-dotted line and open triangle);
% Comet C1 (long-dashed line and open square);
% Comet C/2002~T7 (dash-dotted line and solid triangle);
% Comet Q4 (dashed line and solid square).

The model calculations assume that the ammonia production
rate follows the water production rate with an abundance ratio
of exactly 1.0\%,
a mean lifetime of $\tau \approx$~5600 s (with small corrections for
the phase of the solar cycle),
and an antenna beamwidth of $\theta$ = 41.5$^{\prime \prime}$.
%(66.0$^{\prime \prime}$ for C/1996~B2 (Hyakutake) at the Green Bank Telescope).

The H$_2$O production rate and gas outflow velocity for each comet
were assumed to vary with heliocentric distance $r$ (in AU)
according to (e.g., A'Hearn et al.~1995):

\begin{equation}
Q(r) = Q_o \left[ \frac{r_o}{r} \right]^{\,\alpha} ;  \;\;\;\; 
\delta v(r) = \frac{\delta v_o}{\sqrt{r}}
\label{Q}
\end{equation}
with $\alpha$ (usually $\alpha \approx$~3) and the values of $r_o$,
and $Q_o$ taken from the literature.
 The linewidths at 1~AU  ($\delta v_o \,$) are
converted from the linewidths at each observation epoch ($\delta v \,$) given
in Sec.~\ref{sect:obs}.  
The model parameters and the reference sources used to calculate
the prediction curves of Fig.~\ref{rcomet_log_2004} are listed
in Table~\ref{model_param}.

The unique stature of C/1995~O1 (Hale-Bopp), which had a production rate
much greater than the other comets,
is clearly evident in Fig.~\ref{rcomet_log_2004} and Table~\ref{model_param}.
The solid circle points with rms error bars for C/1995~O1 (Hale-Bopp)
in Fig.~\ref{rcomet_log_2004} are derived
from the measured (3,3) line strengths (Bird et al.~1997)\@.
The water production rate at perihelion, 
1.0 $\times$ 10$^{31}$~s$^{-1}$,
was taken from Colom et al.\ (1999).

The procedure originally developed by Snyder (1982)
is used here to estimate the production rate $Q(S_i)$ of species $i$
from the observed beam-averaged column density $\langle N(S_i) \rangle$
\begin{equation}
\langle N(S_i) \rangle = \frac{4 Q(S_i)}{\pi \; \delta v \; \Delta \theta}
\left\{ \begin{array}{ll}
d/\Delta \theta & {\rm for} \;\; d < \Delta \theta  \\
{\cal F} \left( d/\Delta \theta \right)
                        & {\rm for} \;\; d > \Delta \theta  \\
\end{array}
\right.
\label{QNH3}
\end{equation}
with $d$ the comet coma diameter,
$\Delta$ the geocentric distance,
$\delta v$ the linewidth (taken as 2$u$,
where $u$ is the gas outflow velocity),
and $\theta$ the half-power beam width of the antenna. 
The function $\cal F$, given by
\begin{equation}
{\cal F} = \arccos \left( \frac{ \Delta \theta }{ d } \right) +
\frac{ d }{ \Delta \theta } -
\sqrt{ \left( \frac{ d }{ \Delta \theta } \right)^2 - 1 }
\label{F}
\end{equation}
varies from ${\cal F} = 1$ for $d = \Delta \theta$ to
${\cal F} = \pi/2$ for $d \gg \Delta \theta$.

Using Eq.~(\ref{QNH3}),
we can derive an upper bound on a comet's NH$_3$ production rate from the
upper bounds on the column densities given in the bottom lines
for each comet of Table~\ref{obs_table}.
The resulting NH$_3$ production rates for these comets are given in
Table~\ref{nh3_results} and range between 
1.8 and $23 \times 10^{26}$~s$^{-1}$.

% %
% \hspace*{2mm} $Q(\hbox{NH}_3) \leq 1.4 \times 10^{26}$~s$^{-1}$,
% for Comet C/2001 A2 (LINEAR),\\[1mm]
% %
% \hspace*{2mm}$Q(\hbox{NH}_3) \leq 3.4 \times 10^{26}$~s$^{-1}$,
% for Comet 153P/Ikeya-Zhang,\\[1mm]
% %
% \hspace*{2mm} $Q(\hbox{NH}_3) \leq 6.4 \times 10^{26}$~s$^{-1}$,
% for Comet C/2001 Q4 (NEAT), and\\[1mm]
% %
% \hspace*{2mm}$Q(\hbox{NH}_3) \leq 1.93 \times 10^{27}$~s$^{-1}$,
% for Comet C/2002 T7 (LINEAR).\\[1mm]

\section{Ammonia-to-water ratio}

The strongly variable water production rate for Comet C/2001~A2-B was
reported as $Q(\hbox{H}_2\hbox{O}) = 3.8 \times 10^{28}\hbox{ s}^{-1}$
on 1.7--2.0 July 2001 (Biver et al.~2001), consistent also
  with $Q(\hbox{H}_2\hbox{O}) \sim 4.0\times 10^{28} \hbox{s}^{-1}$
  observed on 9-10~July~2001 (Dello Russo et al.~2005). The Odin
  satellite measured 5--$8\times 10^{28} \hbox{molec. s}^{-1}$ between
  20~June 2001 and 7~July 2001 (Lecacheux et al.~2003).  A recent
  compilation of continuous water production rates derived from SWAS
  observations (Submillimeter Wave Astronomy Satellite) of Comet
  C/2001~A2-B (F.~Bensch, private communication) suggests a value of
  $Q(\hbox{H}_2\hbox{O}) = 2.8 \times 10^{28}$~s$^{-1}$ as appropriate
  for the MPIfR observations on 7~July 2001.  These water production
  rates imply that the maximum NH$_3$ fraction for the Comet
  C/2001~A2-B observations becomes 0.63\%.
  
  An estimate of the water production rate in Comet 153P/Ikeya-Zhang
  of $Q(\hbox{H}_2\hbox{O}) = 1.7 \times 10^{29}$~s$^{-1}$ could be
  derived from Odin satellite observations of the 557 GHz line of
  water on 26.8 April 2002 (Crovisier et al., private communication;
  Lecacheux et al.~2003).  HST ultraviolet OH observations on 20-22
  April 2002 (Weaver et al., private communication) yielded
  $Q(\hbox{H}_2\hbox{O}) = 2.3 \times 10^{29}$~s$^{-1}$.  Newer
  systematic estimates of the H$_2$O production for Comet 153/P
  Ikeya-Zhang (Dello Russo et al.~2004) are smaller by more than a
  factor of two with respect to the original estimates used in the
  NH$_3$ abundance estimates of Bird et al.~(2002).  The best fit
  model prediction of Dello Russo et al.~(2004), the parameters of
  which are given in Table~\ref{model_param}, yields a water
  production rate of $Q(\hbox{H}_2\hbox{O}) = 1.0 \times
  10^{28}$~s$^{-1}$ for the mean epoch of the MPIfR observations.  We
  thus calculate an upper bound on the ammonia-to-water ratio in Comet
  153/P Ikeya-Zhang of 0.63\%.

% new Table 7 from Mike Bird's email of 3 May updating int. times

%%%%%%%%%%%%%%%%%%%%%%%%%%%%%%%%%%%%%%%%%%%%%%%%%%%%%%%%%%%%%%%%%
%                             Table NH3_results
%%%%%%%%%%%%%%%%%%%%%%%%%%%%%%%%%%%%%%%%%%%%%%%%%%%%%%%%%%%%%%%%%
\begin{table*}
\caption[]{NH$_3$ production rates and abundances relative to H$_2$O}
\label{nh3_results}
\vspace*{1mm}
\hspace*{8mm}
\begin{tabular}{lcccl}
\hline
Comet                   &epoch  & $Q(\hbox{NH}_3)$ & $Q(\hbox{H}_2\hbox{O})_r$ &[NH$_3$]/[H$_2$O]\\
            & & [10$^{26}$~s$^{-1}$] & [10$^{29}$~s$^{-1}$]\\ % r &ro \\
\hline
C/2001 A2-B (LINEAR) & 7.22 Jul  &1.8     & 0.28   &$\leq$  0.63\%\\ 
153P/Ikeya-Zhang     &25.42 Apr  &6.4     & 1.0   &$\leq$  0.63\%\\
C/2001 Q4 (NEAT)     & 14.11 May  &2.7     & 2.1   &$\leq$  0.13\%\\ 
C/2002 T7 (LINEAR)   & 11.64 May  &23    & 3.1   &$\leq$  0.74\%\\ 
\hline
\end{tabular}
\end{table*}

Water production rates for the Comets C/2001~Q4 and C/2002~T7
have been reported (N.~Biver, private communication) as 
$Q(\hbox{H}_2\hbox{O}) = 1.8\pm0.2 \times 10^{29}$~s$^{-1}$ on 27.0 April 2004, 
%[Odin satellite]
and 
$Q(\hbox{H}_2\hbox{O}) = 4.5\pm1.0 \times 10^{29}$~s$^{-1}$ on 2.0 May 2004,
respectively.
These lead to upper bounds on the ammonia fractional abundance
relative to water  of, respectively, 0.13\% and 0.74\%.
%Since the column density varies linearly with the 
%production rate, the ratio of the upper bound to the predicted value
%at the same epoch yields the maximum NH$_3$ abundance expressed in percent.  

The upper limits for the relative NH$_3$ abundance
in each comet, based on these H$_2$O production rates and the
NH$_3$ production rates derived in Sec.~\ref{sect:limits}, are
presented in Table~\ref{nh3_results}.
 For comparison,
a summary of the ammonia abundance estimates from all
previous comet observations has been compiled by 
Bockel\'ee-Morvan et al.~(2005). 

% for the last two comets, the NH3 abundance is just Q(NH3)/Qo(H2O),
% but for A2B and Ikeya-Zhang,that doesn't seem to work. Check with Mike.

  All of the upper limits we present here are significantly lower
  than the estimate of 1.1\% for C/1995~O1 (Hale-Bopp) (Bird et
  al.~1999) or 1.5\% for Comet 1P/Halley (Meier et al.~1994).  The
  relative abundance implied for C/2001 Q4 was found to be
  considerably lower than the values of 0.64--0.74\% derived for the other
  three comets.
%remarkably steady
This may be a real effect, but is subject to reconfirmation when
finally agreed values for the water production rate become available.

%Although these are all significantly lower than the estimate of 1.5\%
%for 1P/Halley (Meier et al.~1994) or the published value of 1.1\% for
%C/1995~O1 (Hale-Bopp) (Bird et al.~1999), reevaluation of the NH$_3$
%abundance in C/1995~O1 using revised determinations of
%$Q(\hbox{H}_2\hbox{O})$ suggests an NH$_3$ abundance of 0.7\%, only
%slightly higher than the limits for our four comets (Bockel\'
%ee-Morvan et al.~2005). 
%% NH$_3$ was also detected in Comet Hyakutake
%%(Palmer et al.~1996) with NH$_3$/H$_2$O$ = 0.3$\%.
% [THE HYAKUTAKE DETECTION WAS MENTIONED ALREADY IN THE INTRODUCTION.
% THE REVISED [NH3]/[H2O] WAS 0.6\% (BIRD ET AL. (1999).]

\section{Conclusions}

Our observations suggest that there is a diversity of the NH$_3$
abundance among comets.  This was already suggested by the large (more
than a factor of ten) range of the NH/OH ratio observed from
narrowband photometry in the visible for a large sample of comets by
A'Hearn et al.~(1995).  The nondetections of ammonia in C/2001 A2-B
(LINEAR), C/2001 Q4 (NEAT) and C/2002 T7 (LINEAR), as well as the very
marginal detection in P153 Ikeya-Zhang, lead to the preliminary
indication that the NH$_3$ fraction may be a factor of $\sim$2 lower
than that derived from the detections in Comets 1P/Halley and
C/1995~O1 (Hale-Bopp). 

  With the possible exception of C/2001 Q4 (NEAT), it is interesting
  that the upper limits all seem to be consistent with the 0.5\%
  abundance determined from the NH$_2$ analysis of Kawakita \&
  Watanabe (2002).

The accumulated statistical sample is clearly
insufficient for drawing a final conclusion on the spread in relative
abundances of NH$_3$ in cometary comae.  Nevertheless, the current
comet count does imply that the amount of ammonia in the archetypical
comets 1P/Halley and C/1995~O1 (Hale-Bopp) may be the exception rather
than the rule.  Only more observations will provide final resolution
of the issue.

\begin{acknowledgements}
  
  The results reported in this work are based on observations made
  with the 100-m telescope of the Max-Planck-Institut f\" ur
  Radioastronomie (MPIfR) at Effelsberg.  We are
  grateful to W.J. Altenhoff for carefully preparing the cometary
  ephemerides and to K.M. Menten for discussions and helping to
  arrange for these observations.  We thank N. Biver and F. Bensch for
  providing preliminary water production rates for the comets
  C/2001~Q4 (NEAT)/C/2002~T7 (LINEAR) and C/2001~A2 (LINEAR),
  respectively, and J. Crovisier for his helpful comments as referee,
  and for kindly providing the OH line strengths observed at the
  Nan\c{c}ay RT\@ and the linewidths derived from other molecular
  observations in advance of publication.

\end{acknowledgements}

\end{document}